\documentclass[twocolumn,showpacs,amsmath,amssymb,prb]{revtex4}

\usepackage{graphicx}
\usepackage[usenames]{color}
\usepackage{pstricks}
\usepackage{pst-node}

\newcommand{\GL}{\ensuremath\Gamma_L}
\newcommand{\GR}{\ensuremath\Gamma_R}
\newcommand{\bra}[1]{\ensuremath \left|{#1}\right\rangle}
\newcommand{\ket}[1]{\ensuremath \left\langle{#1}\right|}
\newcommand{\up}{\ensuremath\uparrow}
\newcommand{\dn}{\ensuremath\downarrow}
\newcommand{\x}{\ensuremath e^{-i\chi_{r}}\,}
\newcommand{\X}{\ensuremath e^{ i\chi_{r}}\,}
\newcommand{\CF}{\ensuremath e^{i\chi}}

\newcommand{\cf}{\ensuremath e^{i\frac{\chi}{2}}}

\renewcommand{\vec}[1]{\ensuremath{\mathbf #1}}
\newcommand{\PI}{\boldsymbol{\pi}}
\newcommand{\F}{e^{i\frac{\phi}{4}}}
\newcommand{\f}{e^{-i\frac{\phi}{4}}} 
\newcommand{\Ginf}{\ensuremath \Gamma_{\infty}}
\newcommand{\si}{\ensuremath \sin{\frac{\phi}{2}}}
\newcommand{\co}{\ensuremath \cos{\frac{\phi}{2}}}

\begin{document}

\title{Tunable dynamical channel blockade in double-dot Aharonov-Bohm interferometers}

\author{Daniel Urban}
\affiliation{Theoretische Physik, Universit\"at Duisburg-Essen and CeNIDE, 47048 Duisburg, Germany}

\author{J\"urgen K\"onig}
\affiliation{Theoretische Physik, Universit\"at Duisburg-Essen and CeNIDE, 47048 Duisburg, Germany}

\date{\today}

\begin{abstract}
We study electronic transport through an Aharonov-Bohm interferometer with single-level quantum dots embedded in the two arms. 
The full counting statistics in the shot-noise regime is calculated to first order in the tunnel-coupling strength. 
The interplay of interference and charging energy in the dots leads to a dynamical channel blockade that is tunable by the magnetic flux penetrating the Aharonov-Bohm ring.
We find super-Poissonian behavior with diverging second and higher cumulants when the Aharonov-Bohm flux approaches an integer multiple of the flux quantum. 
\end{abstract}

\maketitle

\section{Introduction}

The study of full counting statistics (FCS) of charge transport through mesoscopic systems has become a well established field. A number of theoretical approaches for calculating the cumulant generating function have been developed,\cite{levitov:1996,QNmesPhys,nazarov:2002,KeldyshGF,stochasticPathint,emary:2007} extended to interacting systems,\cite{TunnelDots,bagrets:2003,kiesslich:2006,gogolin:2006} and memory effects have been included.\cite{braggio:2006,flindt:2008,emary:2009}
Experimentally, it has become possible to count single electrons in real time as they pass through a system of quantum dots.\cite{gustavsson:2006,fujisawa:2006,fricke:2007}
Despite the detector performing a projective measurement, interference has been observed in specially designed systems.\cite{gustavsson:interference}
Apart from the conceptionally straightforward idea of counting individual electrons  there have been numerous proposals for counting statistics detectors, involving qubits and superconducting systems.\cite{qubitdetector,thermalescape,MQTfourthcumulant,thresholddetectors,cooperpairtunneling}

A useful reference to compare transport statistics with are Poissonian processes, which describe uncorrelated events of charge transfer.
Correlations between the transport events change the transport statistics.
For fermionic systems, correlations typically lead to a reduction of the current noise.
However, there are also various scenarios in which the noise is increased.
A prominent example that exhibits super-Poissonian noise is electron bunching, in which periods of high and low (or zero) current alternate.
Bunching behavior has been discussed in complex geometries such as beam splitters\cite{cottet:2004} and serial double\cite{kiesslich:2007,sanchez:2008} and triple\cite{aghassi} quantum dots.
It may be understood as a consequence of a system's bistability.\cite{bistability,jordan:2004} A further example for a bistable system is a quantum shuttle, which exhibits enhanced noise at the transition from the tunneling to the shuttling regime.\cite{flindt:2005}

Super-Poissonian transport behavior can already be found in a single quantum dot, e.g. when lifting of the spin degeneracy of the level results in different tunneling rates for the two spin states.\cite{belzig:2005}
A similar effect was measured in a quantum dot with two states coupling differently to the leads
 either due to their differing spatial extension\cite{gustavsson:2006}
or due to spin-dependent tunneling to ferromagnetic leads.\cite{bulka,braun:noise}
Transport through a single level can also exhibit bunching if a second level in its vicinity interrupts transport by means of Coulomb interaction.\cite{safonov,djuric:2005,kiesslich:2003,gurvitz}
In a similar way, driven transitions between two levels may give rise to enhanced noise.\cite{sanchez:2007}

Also the counting statistics of transport through parallel double quantum dots may exhibit super-Poissonian behavior.
Enhanced noise of the cotunneling current has been found in the presence of ferromagnetic 
leads.\cite{weymann:2008} In DQDs with normal leads enhanced noise was predicted for spinless electrons with\cite{spinless2,spinless1,gurvitz} and without\cite{guo:2008} inter-dot Coulomb interaction and for spinful electrons in double dots with an inter-dot tunnel coupling that leads to a splitting between symmetric and antisymmetric one-electron states in the double dot.\cite{bodoky:2008} 

\begin{figure}[b]
	\includegraphics[width=\columnwidth]{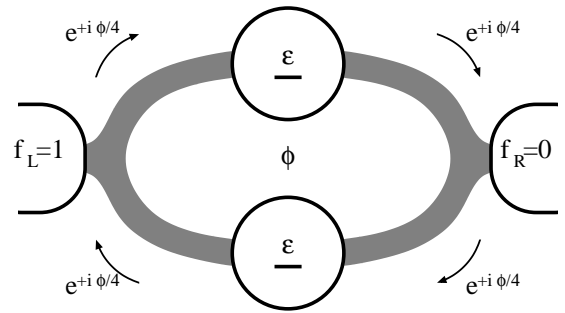}
	\caption{	The double-dot Aharonov-Bohm interferometer consists of two quantum dots ($u$ and $d$) connected to two leads in parallel. The paths through the dots enclose a magnetic flux $\Phi$, giving rise to Aharonov-Bohm interference.}
	\label{fig:system}
\end{figure}

In this paper we study a parallel double quantum dot (DQD) as shown in Fig.~\ref{fig:system}.
The dots are weakly coupled to two leads, operated in the shot-noise regime, with an Aharonov-Bohm flux $\Phi$ enclosed by the two paths. We assume the two dots to be sufficiently separated such that there is no direct tunneling and no inter-dot charging energy. On the other hand, we assume a strong Coulomb interaction within each dot. The system can, therefore, accommodate at most two electrons. 
The spin degree of freedom will turn out to be a vital ingredient, since it necessitates the description of the doubly-occupied states by singlet and triplets. These will turn out to be critical for the decomposition of the system's Liouville space into two disjoint parts, resulting in diverging cumulants.

The possibility to entangle the spins in the dots is used in the context of quantum information 
processing.\cite{burkhard:1999,loss:2000} 
It was recently pointed out\cite{legel:2007,legel:2008} that in non-equilibrium situations
an imbalance between spin singlet and triplet states in the DQD can be generated, based on a scheme that is similar to coherent population trapping.\cite{michaelis,xu}
In contrast to double dots with direct tunneling between the dots, the symmetric and antisymmetric one-electron states remain energetically degenerate for weak tunnel coupling to the leads.
The same holds true for the singlet and triplet two-electron states. The imbalance between singlet and triplet is, therefore, a genuine non-equilibrium effect that relies on the interplay between coherent tunnel coupling and Coulomb interaction.\cite{legel:2007,legel:2008}

In the Aharonov-Bohm interferometer discussed in this paper, the singlet-triplet imbalance depends on the Aharonov-Bohm phase, which can be tuned by the magnetic flux enclosed by the interferometer arms. Its origin is related to the fact that the source and drain electrons only couple to certain linear combinations of the upper and lower dot levels. Most dramatic effects are expected for Aharonov-Bohm fluxes that are close to integer multiples of the flux quantum. In this case, as we will discuss in more detail below, the singlet and triplet states belong to two different subspaces of the double dot's Hilbert space.
These subspaces are nearly disconnected from each other and are described by different transport characteristics.
As a consequence, we will find bunching behavior that leads to not only an enhancement but even a divergence of the second and higher cumulants as a function of the Aharonov-Bohm phase.

The paper is structured as follows: In Section~\ref{sec:system} we specify the system model. Section~\ref{sec:meq} describes how to calculate the full counting statistics. The results are discussed in Sec.~\ref{sec:results}, where we illustrate the mechanism of super-Poissonian statistics and discuss several limiting cases. In Sec.~\ref{sec:spinless} we demonstrate the origin of the effect by comparing to a spinless model. Finally, we conclude in Sec.~\ref{sec:conclusion}.

\section{System}	\label{sec:system}

The double-dot interferometer shown in Fig.~\ref{fig:system} is described by the Hamiltonian
\begin{equation}\label{eq:hamiltonian}
   H = H_u+H_d + H_L+H_R +H_{T}	\, .
\end{equation}
The quantum dots, $H_i = \sum_{\sigma} \varepsilon_{i}\, c_{\sigma i}^\dagger
c_{\sigma i} + U n_{\up  i}n_{\dn i}$ for $i=u,d$, are described as Anderson impurities
with spin-degenerate electronic levels $\epsilon_i$ and charging energy $U$ for double occupation. 
Throughout this paper we are interested in the regime of strong Coulomb interaction ($U$ greater than all other energies), so that at most single occupation of each dot is allowed.
Furthermore, we concentrate on the situation when both quantum-dot levels are tuned close to each other. We define the average level energy as $\epsilon = (\epsilon_{u} + \epsilon_{d})/2$.
Each of the leads is described as a reservoir of non-interacting fermions $ H_r = \sum_{k\alpha\sigma} \varepsilon_{rk\sigma}^{}\, a_{rk\sigma}^\dagger a_{rk\sigma} $ with indices for lead $r \in \{L,R\}$, momentum $k$ and spin $\sigma$.
The tunneling Hamiltonian $H_{T} = \sum_{r,i} H_{T,ri}$ consists of parts  for tunneling between each dot $i$ and each lead $r$,
\begin{equation}
	H_{T,ri}	=	\sum_{k,\sigma}     \, t_{ri}   a^\dag_{rk\sigma} c_{\sigma i}^{} + \text{H.c.}        ,
\end{equation}
with the flux dependence included in the phases of the tunneling amplitudes
$t_{L,d} = |t_L| e^{i\phi/4}$,
$t_{L,u} = |t_L| e^{-i\phi/4}$,
$t_{R,u} = |t_R| e^{i\phi/4}$, and
$t_{R,d} = |t_R| e^{-i\phi/4}$, according to Fig.~\ref{fig:system}. The phase $\phi$ is related to the magnetic flux $\Phi$ through the ring as $\phi=2\pi\Phi/\Phi_{0}$, where $\Phi_{0}= h/e$ is the flux quantum.
The tunneling rate through interface~$r$ is quantified by $ \Gamma_r/\hbar =  2 \pi \left|t_{r}\right|^2 \rho_{r}/\hbar $. For simplicity, we assume the density of states $\rho_r$ and the tunneling amplitudes $t_r$ to be independent of energy, which implies constant tunneling rates.

In addition to strong on-site Coulomb repulsion, we assume no interdot interaction, so that the entire system can be occupied by at most two electrons.
The probabilities to find the system empty and singly occupied are $p_{0}$ and $p_{1}$, respectively.
Charging the empty system with an electron of spin $\sigma$ from the left lead results in the state 
$( \F c^{\dag}_{u\sigma} + \f c^{\dag}_{d\sigma} )/\sqrt{2} \bra{0}  =  (\F\bra{\sigma,0}+\f\bra{0,\sigma})/\sqrt{2} \equiv \bra{+}_L$. This state is not fully described by the probability of single occupation $p_{1}=p_{u}+p_{d}$, it rather needs to be further specified by off-diagonal elements of the density matrix
$p_{\nu}^{\mu} = \bigl\langle \bra{\nu}\ket{\mu} \bigr\rangle$ where $\mu$ and $\nu$ label the nine dot states
$ \bra{0}, \bra{\up,0}, \bra{\dn,0}, \bra{0,\up}, \bra{0,\dn}, \bra{\up,\up}, \bra{\dn,\dn}, \bra{\up,\dn}, \bra{\dn,\up} $.
We summarize them as an isospin in the two-dimensional Hilbert-space of the two dot levels 
$\vec{I}_{\sigma} = (I_{\sigma,x},I_{\sigma,y},I_{\sigma,z}) = ( p^{u}_{d}+p^{d}_{u} , i(p^{u}_{d}-p^{d}_{u}) , p_{u}-p_{d} )/2$. 
In this basis the state reached by tunneling in from the left lead is isospin-polarized along $\vec{n}_{L}=(\cos{\phi/2},\sin{\phi/2},0)$. The right lead is correspondingly isospin-polarized along $\vec{n}_{R}=(\cos{\phi/2},-\sin{\phi/2},0)$.

Due to strong Coulomb interaction on the dots double occupation of the system is allowed only if one electron is found in each dot.
This means that by sequential
filling from the source lead only the singlet state $\bra{S}=(\bra{\up,\dn}-\bra{\dn,\up})/\sqrt{2}$ is accessible. The three triplets $\bra{T_{+}}=\bra{\up,\up}$, $\bra{T_{-}}=\bra{\dn,\dn}$, $\bra{T_{0}}=(\bra{\up,\dn}+\bra{\dn,\up})/\sqrt{2}$ can be accessed only indirectly as we will see later. Spin symmetry of the Hamiltonian requires all triplets to be occupied with equal probability $p_{T_{+}}=p_{T_{-}}=p_{T_{0}}=p_{T}$ in the stationary limit.
While in principle the density matrix may contain $9\times9=81$ elements, this number reduces to the following seven:
$p_{0},p_{1},p_{S},p_{T}$ and three isospin components $\vec{I}$ (spin symmetry requires the occupation of all triplets to be equal).

\section{Master Equation and FCS}	\label{sec:meq}

The time evolution of the system's density matrix can be described by a number-resolved generalized master equation
\begin{eqnarray}\label{eq:meq}
	&&\frac{d}{dt} p_{\nu}^{\mu} (N,t)	\,+\,	i (\varepsilon_\mu - \varepsilon_\nu) \, p_\nu^\mu(N,t)	\\ \nonumber
	&&	=	\sum_{N'=-\infty}^{\infty} \int_0^{t} dt' \sum_{\mu',\nu'}\,
	W_{\nu\nu'}^{\mu\mu'}(N-N',t-t') \, p_{\nu'}^{\mu'}(N',t')	,
\end{eqnarray}
where $p_{\nu}^{\mu} (N,t)$ is the component of the density matrix under the condition that $N$ electrons have passed the system after time $t$. In general, the energy difference $(\varepsilon_\mu - \varepsilon_\nu)$ appears in the presence of off-diagonal matrix elements. It, however, vanishes for symmetric dot level energies, $\epsilon_{u}=\epsilon_{d}$.

The kernel of this equation can be obtained using a diagrammatic real-time technique formulated on the Keldysh contour. It allows for a perturbative expansion in the coupling strength, which we abort in lowest order $\Gamma$ to describe the weak-tunneling limit. For a detailed derivation of this diagrammatic language and its rules for the calculation of diagrams we refer to Refs.~\onlinecite{diagrams,technique,legel:2007,legel:2008}. Not described in these references is the inclusion of the counting field $\chi$, the Fourier-conjugated variable of the transferred charge $N$. It is introduced at each junction by replacing the tunnel matrix elements in the Hamiltonian as $t_{r} \rightarrow t_{r} e^{\pm i\chi_{r}/2}$ with $\chi_{L}=-\chi_{R}=\chi/2$, where the positive (negative) sign is taken for vertices on the upper (lower) branch of the Keldysh contour.
The counting field is thus attached to both interfaces in such a way that only electrons passing the entire system contribute to the statistics.

The cumulant generating function is defined as
\begin{equation}\label{eq:S}
	S(\chi,t_0)	=	\ln \left[ \sum_{N=-\infty}^{\infty} e^{i\chi N} P(N,t_0) \right]
\end{equation}
where $P(N,t_{0})=\sum_{\mu} p_{\mu}^{\mu} (N,t_{0})$ labels the probability that $N$ electrons have passed the system and $p_{\mu}^{\mu}$ are the diagonal elements of the systems density matrix, i.e.~the occupation probability of state $\bra{\mu}$.
The cumulants can be obtained by taking derivatives with respect to the counting field
$\kappa_{n}=(-i)^{n}(e^{n}/t_{0}) \partial^{n}/\partial\chi^{n}S(\chi)$.
The density matrix elements $p_{\mu}^{\nu}$ are the solutions of the master equation Eq.~(\ref{eq:meq}) in the steady state.
In order to solve Eq.~(\ref{eq:meq}), we first Fourier-transform the equation with respect to $N$, thereby introducing the counting field $\chi$ and second perform a Laplace transform in time.
We obtain the cumulant generating function by following the same steps as in Ref.~\onlinecite{braggio:2006}, in which, however, only generalized master equations for a diagonal density matrix were considered.
This is, e.g., sufficient to describe Aharonov-Bohm interferometers that contain one quantum dot.\cite{urban:2008} 
In order to describe a double-dot Aharonov-Bohm interferometer, however, we need to extend the approach of Ref.~\onlinecite{braggio:2006} to include also non-diagonal density matrix elements.

It turns out that this extension is quite straightforward. All the formal steps of Ref.~\onlinecite{braggio:2006} remain the same. The only difference is that when writing the generalized master equation in a matrix notation, off-diagonal matrix elements need to be taken into account as well, i.e., 
$\dot{\PI}(\chi)=\vec{W}(\chi)\cdot\PI(\chi)$ where $\PI=(p_{\mu_{1}}^{\mu_{1}},p_{\mu_{2}}^{\mu_{2}},\ldots,p_{\mu_{n}}^{\mu_{n}}  ,  p_{\rho_{1}}^{\sigma_{1}},p_{\rho_{2}}^{\sigma_{2}},\ldots,p_{\rho_{m}}^{\sigma_{m}})$ first collects all diagonal and then all off-diagonal matrix elements of the density matrix
(the indices $\mu_i,\rho_i,\sigma_i$ label the system states.).
As a consequence, we need to redefine the vector $\vec{e}^{T} = (1,\ldots,1,0,\ldots,0)$, which is needed to compute $P(N)=\vec{e}^{T}\cdot\PI(N)$.
The central result of Ref.~\onlinecite{braggio:2006} is still valid, namely Eq.~(5) and its generalization to non-Markovian orders. It relates the cumulant generating function to the eigenvalue $\lambda(\chi)$ of the kernel $\vec{W}(\chi)$, whose real part has the smallest absolute (negative) value.
The only modification is that the matrix $\vec{W}(\chi)$ is enlarged since it allows for transitions from and to off-diagonal states as well. 
In lowest order in the tunnel coupling strength, the cumulant generating function is simply given by
\begin{eqnarray}\label{eq:Smarkov}
	S(\chi,t_0)	=	t_{0} \, \lambda(\chi)	.
\end{eqnarray}

In many cases the matrix $\vec{W}$ is too large and complex to obtain its eigenvalues analytically.
In such situations, one can nevertheless obtain all the cumulants recursively. This has been demonstrated for master equations similar to Eq.~(\ref{eq:meq}) by Flindt {\it et al}.\cite{flindt:2008}
This technique can be applied to systems with non-diagonal density matrix elements without modification. Details and analytic forms for the the first cumulants can be found in the appendix and the literature.

We present the master equation for the case of degenerate dot levels (i.e.~level splitting $\Delta\epsilon=0$) using an intuitive notation distinguishing between occupation probabilities $\vec{p}=(p_{0},p_{1},p_{S},p_{T})$ and isospin $\vec{I}$
 [in the above notation $\PI=(\vec{p},\vec{I})$].
\begin{widetext}
\begin{equation}	\label{eq:transitionrates}
	\frac{d}{dt} \vec{p}	=	\sum_{r=L,R}
					\Gamma_{r} \left(\begin{array}{cccc}
					-4 f_r	& \x (1-f_r)		& 0		& 0		\\
					\X 4 f_r& -(1+f_r)		& \x 2(1-f_r)	& \x 2(1-f_r)	\\
					0	& \X \frac{1}{2} f_r	& -2(1-f_r)	& 0		\\
					0	& \X \frac{3}{2} f_r	& 0		& -2(1-f_r)	
				\end{array}\right) \cdot \vec{p}
					\; + \;
				\Gamma_{r} \left(\begin{array}{c}
					\x 2(1-f_r)	\\	-2(1-2f_r)	\\	\X  f_r	\\	-\X 3f_r	
				\end{array}\right) \vec{I}\cdot\vec{n}_{r}
\end{equation}
\begin{equation}
	\frac{d}{dt} \vec{I}	=	\sum_{r=L,R}
					\Gamma_{r} \left[  \X 2f_{r} p_{0}  +  (1-f_{r}) p_{1}  +  \x(1-f_{r}) p_{S}  -  \x(1-f_{r}) p_{T}  \right]  \; \vec{n}_{r}
					- \Gamma_{r} (1+f_{r}) \vec{I}
\end{equation}
\end{widetext}
In the following we will only be interested in the shot-noise regime $eV \gg k_BT$. In particular, we consider the situation that the dot levels lie inside the energy window defined by the Fermi levels of the leads, so that $f_{L}(\epsilon)=1$, $f_{R}(\epsilon)=0$.

\section{Flux Dependent Counting Statistics}	\label{sec:results}

To illustrate the origin of the singlet-triplet imbalance discussed in Ref.~\onlinecite{legel:2007}, we transform the basis of the double-dot states with a transformation matrix $\vec{S}$ such that 
$\vec{S}\,\vec{\PI} = (p_{0},\frac{p_{1}}{2}+\vec{I}\cdot\vec{n}_{L},p_{S} , \frac{p_{1}}{2}-\vec{I}\cdot\vec{n}_{L},p_{T} , \vec{I}\cdot(\vec{n}_{z}\times\vec{n}_{L}) , I_{z})$.
The first three elements correspond to the double dot being empty $|0\rangle$,
singly occupied in the symmetric state $\bra{+}_L= (e^{i\phi/4}\bra{\sigma,0}+e^{-i\phi/4}\bra{0,\sigma})/\sqrt{2}$, that is reached by tunneling in from the left lead, and doubly occupied with a spin singlet $|S\rangle$.
In the following this set of states is referred to as the $+$ subspace, named after the isospin in the singly occupied state.
This basis choice is motivated by the fact that for flux $\phi=2\pi m$, the $+$ subspace is not connected by tunneling to the remaining states.
These other states can again be divided into two uncoupled subspaces. The first one consists of single occupation in the state
$|-\rangle_R = ( e^{-i\phi/4}\bra{\sigma,0} - e^{i\phi/4}\bra{0,\sigma})/\sqrt{2}$
(this becomes the antisymmetric state for $\phi=2\pi m$),
and double occupation with a spin triplet $|T\rangle$. The set of these two states is referred to as the $-$ subspace. The remaining components of the isospin that are orthogonal to $\vec{n}_{L}$ are called the $\perp$-subspace. In the case $\phi=0$ these states are not populated and the system dynamics is governed entirely by the $+$- and $-$ subspaces.

The basis change includes no approximation and contains the same information found in Eq.~\ref{eq:transitionrates}, since states and rates were transformed together.
In the new basis the master equation assumes block-diagonal form for $\phi=0$:
\begin{widetext}
\begin{equation}\label{eq:blockdiagonal}
	\vec{S}\vec{W}\vec{S}^{-1} =
	\left(\begin{array}{ccc|cc|cc}
-4\GL		& \cf 2 G_R^+		& 0		& \cf 2 G_R^-		& 0		& \cf 2\GR\si	& 0	\\
\cf 4 G_L^+	& -\GL-G_L^--2G_R^+ 	& \cf 2 G_R^+	& 0			& \cf 2 G_R^-	& -\Gamma \si	& 0	\\
0		& \cf G_L^+		& -2\GR		& \cf G_L^-		& 0		& -\cf \GL \si	& 0	\\\hline
\cf 4 G_L^-	& 0			& \cf 2 G_R^-	& -\GL-G_L^+-G_R^-	& \cf 2 G_R^+	& -\Gamma \si	& 0	\\
0		& \cf 3 G_L^-		& 0		& \cf 3 G_L^+		& -2 \GR	& \cf 3 \GL\si	& 0	\\\hline
-\cf 4 \GL \si	& -\frac{1}{2}\Gamma\si	& \cf \GR \si	& -\frac{1}{2}\Gamma\si	& -\cf \GR \si	& -2\GL-\GR	& 0	\\
0		& 0			& 0		& 0			& 0		& 0		& -2\GL-\GR	
	\end{array}\right)
\end{equation}\end{widetext}
where the following definitions were used: $G_{r}^{\pm}=\Gamma_{r}\frac{1}{2}(1\pm\cos{\frac{\phi}{2}})$, $\Gamma=\GL+\GR$ and $\Ginf=2\GL+\GR$. Note that for $\phi=0$ the matrix assumes block-diagonal form, since $G_r^+ \rightarrow \Gamma_{r}$ and $G_{r}^{-} \rightarrow 0$.

It turns out that for zero flux the master equation assumes block diagonal form and the $+$- and $-$ subspaces decouple.
The $+$ subspace is found in the upper-left $3\times 3$ block.
In presence of a flux it is coupled by tunneling, i.e.~a change in the charge state, to the $2\times 2$-dimensional $-$ subspace located in the middle. 
This coupling cannot be described simply by a rate $\Omega_{+\leftrightarrow-}$.
Instead there are six possible transition paths which occur with four different rates.
The paths and rates can be read off from the $2\times3$- and $3\times2$-blocks in Eq.~\eqref{eq:blockdiagonal} that contain factors $G_r^\pm$.
The two rightmost columns and two lowest lines of Eq.~\ref{eq:blockdiagonal} describe the intermediate occupation of the $\perp$-subspace occurring for $\phi\neq 2\pi m$.

\begin{figure}
	\includegraphics[width=.75\columnwidth]{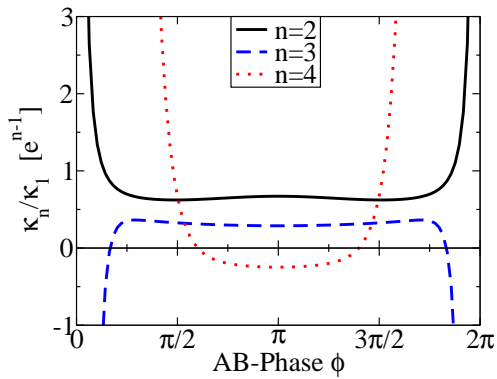}
	\caption{(Color online) The normalized $n$th cumulants $\kappa_{n}/\kappa_{1}$ diverge for $\phi\rightarrow 2\pi m$ in a symmetric system ($\Gamma_{L}=\Gamma_{R}$), due to the competition of two channels belonging to different system states. The width of the divergence is governed by the relaxation rate between these two states.}
	\label{fig:symmetriccumulants}
\end{figure}

As for $\phi=2\pi m$ the system decomposes into two uncoupled subsystems it is no longer possible to calculate its counting statistics as described in Ref.~\onlinecite{braggio:2006}. There are two independent stationary solutions of the master equation~(\ref{eq:meq})--one for each subspace. This means that the (long time) counting statistics would unphysically depend on the initial condition. In realistic systems some kind of coupling would always be present, lifting the degeneracy. Thus we study only small but finite values of the flux $\phi\approx 2\pi m$ so that the full counting statistics is well defined.

\subsection{Channel Exclusion for $\phi=2\pi m$}
\begin{figure}[b]
	\includegraphics[width=.95\columnwidth]{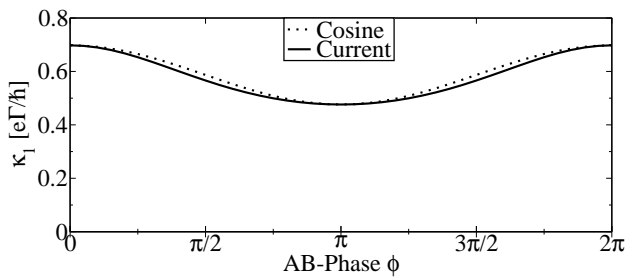}
	\caption{The current through the double dot system is subject to Aharonov-Bohm oscillations. These are almost cosine-like so that the current remains finite for $\phi=0$.}
	\label{fig:current}
\end{figure}

The separation of the system's Hilbert space into two separate subspaces has consequences for the transport statistics:
As discussed in the introduction, systems with several states differing in average current exhibit bunching.
The larger the current difference is, the more enhanced is the noise.
As the rate with which the system switches between the states is decreased, the noise is expected to be enhanced further.
In the system discussed, the coupling can be decreased to zero by tuning the magnetic flux to $\phi=2\pi m$. In this case, the subspaces decouple and all normalized cumulants diverge as the effective charge goes to infinity (see Fig.~\ref{fig:symmetriccumulants}).
\cite{note-1}
We emphasize that the divergence of the normalized cumulants is not caused by a vanishing current, whose dependence on flux is approximately cosinelike (see Fig.~\ref{fig:current}).

To specify the picture outlined above we analyze the properties of the subspaces separately. For this purpose we calculate the cumulant generating functions associated with the $3\times3$- and $2\times2$-subspaces for singlet and triplet, respectively. These are defined as the eigenvalues of the corresponding sub-matrices of Eq.~(\ref{eq:blockdiagonal}).

The cumulant generating function for the $-$ subspace, Eq.~(\ref{eq:ST}) is that of a two-state system,
\begin{equation}	\label{eq:ST}
	S_{-}	=	-\frac{3\GL+2\GR}{2}\left( 1-\sqrt{ 1 + \frac{4(3\GL)(2\GR)}{(3\GL+2\GR)^2}(e^{i\chi}-1) } \right)  .
\end{equation}
The transition rates are $3\GL$ for filling and $2\GR$ for emptying the double dot. This can be understood by counting the possible realizations of each state: If the dot is in a triplet state, taking away either of the two electrons results in single occupation. On the other hand, starting from a singly occupied state there are three triplets which can be accessed by tunneling into the system.

The cumulant generating function for the $+$ subspace is too complex to be shown here.
It describes a three-state system, with transition rates as can be read off from Eq.~(\ref{eq:blockdiagonal}). 
Below, we will give compact analytic expressions in the limit of very asymmetric tunnel couplings to the left and right lead.

The complex internal dynamics distinguish our system from others in the literature in two ways: First, the two states are not just differing in current, but are each characterized by their own distribution function.
Second, the transitions between the subsystems cannot just be described by simple rates $\Gamma_{+ \leftrightarrow -}$. 
Instead there are six possible transition paths which occur with four different rates. These paths and rates can be found in the $2\times3$- and $3\times2$-submatrices in Eq.~\eqref{eq:blockdiagonal}.

Our work also differs from other studies of the counting statistics related to singlet and triplet states\cite{hassler:2008,taddei:2002} in so far as these studied the different statistics of two-particle states propagating along a device, while in our case only a single particle propagates at a time.

\subsubsection{Asymmetric Tunnel Coupling}

\begin{figure}
	\includegraphics[width=\columnwidth]{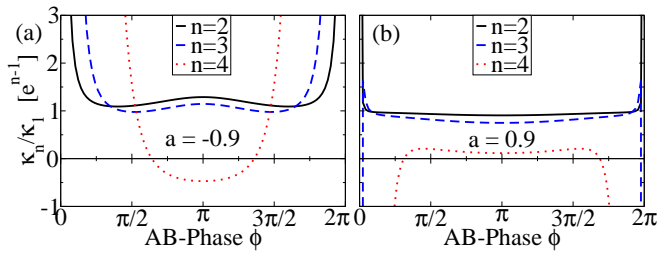}
	\caption{(Color online) The normalized $n$th cumulants $\kappa_{n}/\kappa_{1}$ of the non-symmetric system diverge (left plot: $a=(\GL-\GR)/(\GL+\GR)=-0.9$, right plot: $a=+0.9$). The width of the divergence is slightly enhanced $a\rightarrow -1$ but strongly reduced for $a\rightarrow +1$.}
	\label{fig:asymmetriccumulants}
\end{figure}

In Fig.~\ref{fig:asymmetriccumulants}, we show the second, third, and fourth normalized cumulant for 
asymmetric tunnel coupling to source and drain, parametrized by the asymmetry parameter
$a=(\GL-\GR)/\Gamma$, where $\Gamma=\GL+\GR$ denotes the total coupling.
For $a\rightarrow -1$, the bottleneck for transport is the tunnel barrier between source electrode and double dot.
We find that the width of the divergence as a function of the Aharonov-Bohm flux is slightly increased.
Furthermore, the divergence of the third cumulant has changed its sign as compared to the case of symmetric coupling.
The cumulant generating functions for the $+$- and $-$ subspace at $\phi=2\pi m$,
\begin{eqnarray}
	\left. S_{+} \right|_{a\rightarrow-1}	&=&	4\GL(\CF-1)	\nonumber\\& & \, - \,	\frac{3}{2}\Gamma(a+1)^2 \CF(\CF-1), \\
	\left. S_{-} \right|_{a\rightarrow-1}	&=&	3\GL(\CF-1)	\nonumber\\& & \, - \,	\frac{9}{8}\Gamma(a+1)^2 \CF(\CF-1),
\end{eqnarray}
become Poissonian for $a\rightarrow -1$, with different tunneling rates for the $+$-and $-$ subspace. 
In addition, the probability to find the double dot in a $+$ subspace is higher than that for the $-$ subspace,
$P_{+}=p_0+p_++p_S=\frac{4}{5}+\mathcal{O}(a^{1})$ and $P_{-}=p_-+p_T=\frac{1}{5}+\mathcal{O}(a^{1})$.
These probabilities were obtained from the probabilities calculated in the non-degenerate case $\phi\neq 0$ and then taking the limit $\phi\rightarrow 0$.

\begin{figure}
	\includegraphics[width=.95\columnwidth]{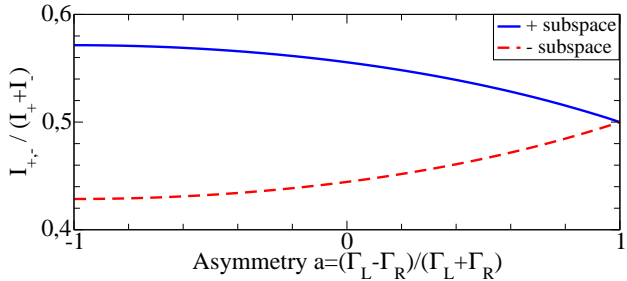}
	\caption{(Color online) For stronger coupling to the drain ($a<0$) the current associated with the $+$ subspace is larger that that for the $-$ subspace. Additionally, the occupations of the subspaces differ. This results in super-Poissonian statistics. For $a\rightarrow1$ both currents and occupations are equal. This yields Poissonian statistics.}
	\label{fig:occcur}
\end{figure}

In the opposite limit, $a\rightarrow 1$, the width of the divergence is strongly suppressed.
In this case, tunneling out of the double dot to the drain electrode defines the bottleneck of transport.
At first glance a stronger bunching may be expected, since with decreasing tunnel coupling to the drain the coupling between the $+$-and $-$ subspaces is reduced.
However, it turns out that in the limit $a\rightarrow +1$, the cumulant generating functions
\begin{eqnarray}
	\left. S_{+} \right|_{a\rightarrow1}	&=&	2\GR(\CF-1)	\nonumber\\&&\, - \,	\frac{1}{4} \Gamma(a-1)^3 e^{2i\chi}(\CF-1)	\\
	\left. S_{-} \right|_{a\rightarrow1}	&=&	2\GR(\CF-1)	\nonumber\\&&\, - \,	\frac{1}{3} \Gamma(a-1)^2 \CF(\CF-1)		
\end{eqnarray}
become identical.
Therefore, the transport statistics is the same for both subspaces, and bunching does not appear anymore.
Figure~\ref{fig:occcur} shows the current $I_{+}$ ($I_{-}$) under the condition that the system is in the $+$ ($-$) subspace as a function of the the asymmetry.
These currents are obtained from the cumulant generating functions for each subspace.
They become equal for $a\rightarrow+1$.

\subsubsection{The influence of relaxation}

\begin{figure}
	\includegraphics[width=.75\columnwidth]{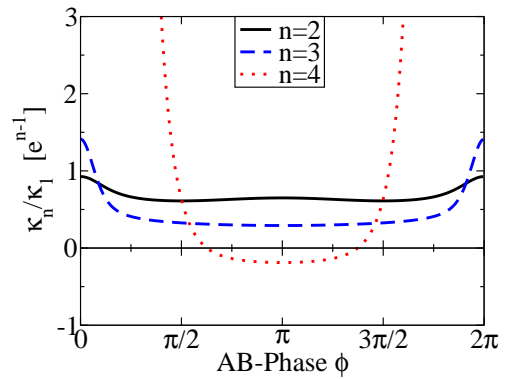}
	\caption{(Color online) Normalized cumulants in the presence of isotropic isospin relaxation (rate $\Omega_{\vec{I}}=\Gamma/10$). The divergent cumulants are suppressed: while the Fano-factor is sub-Poissonian for the relaxation rate shown, higher cumulants still show super-Poissonian behavior.}
	\label{fig:Irel}
\end{figure}

In experiments interaction with the environment can be expected to induce various relaxation mechanisms. Relaxation of the isospin, mediated by electric interactions, will be of particular importance. We model it by introducing a relaxation term with rate $\Omega_{\vec{I}}$ in the master equation Eq.~(\ref{eq:transitionrates}) that reduces the isospin isotropically:
\begin{equation}
	\left(\frac{d}{dt} \vec{I}\right)_\text{rel}	=	-\Omega_{\vec{I}} \, \vec{I}  .
\end{equation}
The effect of this relaxation is primarily a reduction of the visibility of the AB-signal, due to the electrons loosing their coherence. Furthermore, it leads to an effective coupling of the $+$- and $-$ subspaces. Correspondingly the bunching effect is weakened, resulting in the cumulants assuming finite values also for $\phi=2\pi m$ (see Fig.~\ref{fig:Irel}). The figure shows the situation for $\Omega_{\vec{I}}=\Gamma/10$. For sufficiently fast relaxation all cumulants become sub-Poissonian, but as can be seen from the figure, the higher the moment, the faster is the required relaxation rate.

Another relaxation mechanism may be given by spin-flip processes converting singlets into triplets and vice versa. Since we summarized all the triplet occupations in $p_{T}$, terms have to be added to the master equation in the following way:
\begin{eqnarray}
	\left(\frac{d}{dt}p_{S}\right)_\text{rel}	&=&	-\Omega_{ST}	\, p_{S}	+ 3\Omega_{ST} \, p_{T} 	\\
	\left(\frac{d}{dt}p_{T}\right)_\text{rel}	&=&	+\Omega_{ST}	\, p_{S}	- 3\Omega_{ST} \, p_{T}	.
\end{eqnarray}
The factor of three is required to take into account that the triplet probability corresponds to three states, while there is only one singlet.

\begin{figure}
	\includegraphics[width=.75\columnwidth]{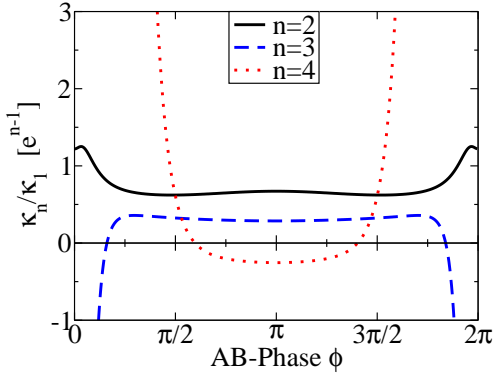}
	\caption{(Color online) Normalized cumulants in the presence of singlet-triplet relaxation (rate $\Omega_{ST}=\Gamma/200$). The influence of the S--T relaxation is much stronger than that of the isospin relaxation.}
	\label{fig:STrel}
\end{figure}

The $+$- and $-$ subspaces are now directly coupled and the divergencies vanish more rapidly as a function of the relaxation rate than in the case of isospin relaxation (Fig.~\ref{fig:STrel}). However, since such a relaxation is mediated magnetically, it can be expected to be much slower than isospin relaxation.

\begin{figure}
	\includegraphics[width=.75\columnwidth]{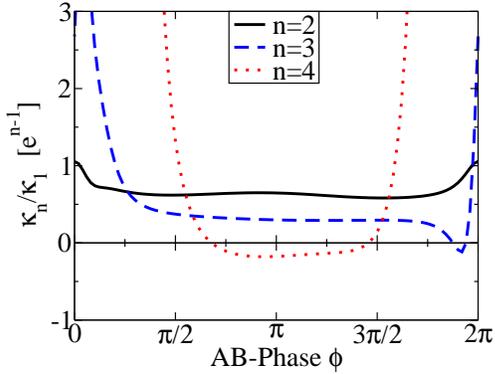}
	\caption{(Color online) Normalized cumulants in the presence of a finite level detuning $\Delta\epsilon=1/3\Gamma$. Symmetry of the flux-dependence about $\phi=\pi$ is lost, but enhancement of the cumulants are still clearly visible.}
	\label{fig:Iprec}
\end{figure}

A third way of coupling the $+$- and $-$ subspaces is not related to relaxation. As can be seen from the master equation, Eq.~\eqref{eq:meq}, a detuning of the energy levels $\Delta\epsilon=\epsilon_{u}-\epsilon_{d}$ gives rise to additional terms. It turns out that for small detuning, $\Delta\epsilon\sim\Gamma$, this results in precession of the isospin about the axis $\vec{n}=(0,0,1)$,
\begin{equation}
	\left(\frac{d}{dt}\vec{I}\right)_\text{prec}	=	\Delta\epsilon \; \vec{n}\times\vec{I}  .
\end{equation}
Since three spatial directions ($\vec{n}_{L}$, $\vec{n}_{R}$ and $\vec{n}$) appear in the master equation, the symmetry of the flux-dependence about $\phi=\pi$ is lost. In other words, the statistics depends on the direction of transport even in the case of symmetric coupling $\GL=\GR$. A similar effect was predicted for a quantum dot with three ferromagnetic leads.\cite{urban:sftrans} For reasonably large values of the detuning, $\Delta\epsilon=\Gamma/3$, enhancement of the moments clearly persists (Fig.~\ref{fig:Iprec}).

\subsection{Super-Poissonian statistics for $\phi=(2n+1)\pi $}

\begin{figure}
	\includegraphics[width=.55\columnwidth]{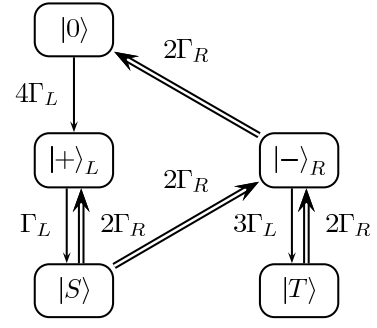}
    \caption{Internal system dynamics for flux $\phi=\pi$ and strong coupling to the drain $a\rightarrow -1$. The double lines denote transitions much faster (with rate $2\GR$) than the single lines (with rates $\propto\GL$).}
    \label{fig:phipi}
\end{figure}

Figure~\ref{fig:asymmetriccumulants} reveals that the statistics for $\phi=(2n+1)\pi$ are also peculiar for $a\rightarrow -1$: The noise and the third normalized cumulant are enhanced beyond the Poissonian value, while the fourth normalized cumulant remains negative. The enhanced noise can be understood by studying the states which predominantly contribute to transport. They can be read off from the master equation Eq.~(\ref{eq:meq}) and are summarized in Figure~\ref{fig:phipi}.

Due to the strongly different coupling strengths, filling the dot is much slower than emptying. Therefore one could expect that the double dot is predominantly empty, while single and double occupation are strongly suppressed. However, the isospin of an electron originating from the left lead has no overlap with the isospin polarization of the right lead. The direct transition $\bra{+}_L\rightarrow\bra{0}$ is therefore forbidden and single occupation turns out to be more likely than an empty dot, as can be seen from the stationary occupation probabilities for $a\approx-1$:
\begin{eqnarray}
	\vec{p}	&\approx&	\frac{1}{5}
					\left(\begin{array}{c}
					1 - (a+1)	\\	4	\\	a+1	\\0
					\end{array}\right)	+\mathcal{O}((a+1)^2)
					\\
	\vec{I}	&\approx&	\left(\begin{array}{c}
					0	\\	-\frac{2}{5}+\frac{a+1}{5} \\ 0
					\end{array}\right)	+\mathcal{O}((a+1)^2)
					,
\end{eqnarray}
with the isospin being parallel to $\vec{n}_L$.
Due to the fact that filling the dot with a single electron occurs with rate $4\GL$ and adding a second electron only with rate $\GL$ the singly occupied state $\bra{+}_L$ is, in lowest order in $(a+1)$, four times more likely than an empty dot.
Occupation of the triplet is even rarer than singlet occupation: it starts in order $(a+1)^{2}$, because it can only be reached via singlet occupation and subsequent decay to the state $\bra{-}_{R}$.

It is eminent from the flowchart Fig.~\ref{fig:phipi} that there are several distinct cycles through which electrons are transported from left to right: the transitions $\bra{+}_{L}\leftrightarrow\bra{S}$ and $\bra{0}\rightarrow\bra{+}_{L}\rightarrow\bra{S}\rightarrow\bra{-}_{R}$ and then back to $\bra{0}$, or several sub-cycles via $\bra{T}$.
As these cycles transfer electrons at different mean currents and with different statistics, it is clear that a complicated telegraph effect will lead to increased noise.
In contrast to the channel exclusion described in the previous sections this effect is not related to separated Hilbert spaces.

\section{Importance of spin}	\label{sec:spinless}

We would like to remark that the divergence of the normalized cumulants for $\phi=2\pi m$ depends crucially on the inclusion of spin in the description of the system. The literature knows a number of examples where finite noise was found in similar, although not equal, double dot systems with spinless electrons.\cite{spinless1,spinless2,guo:2008}
There is also one example where infinite noise is predicted, although spin was not included.\cite{gurvitz} At the end of this section we will discuss the relation to the present model.

Neglecting spin in our model reduces the dimensionality of the Hilbert space to four (instead of nine) since the distinction between singlet and triplet becomes impossible. Instead there is only one doubly-occupied state, which is sufficiently described by its occupation probability $p_{2}$. There are thus in principle $16$ independent density matrix elements (instead of $81$) of which only $6$ (instead of $7$) are independent.
Again, we arrange the density matrix elements in a vector $\PI=(p_{0},p_{1},p_{2},\vec{I})$, so that the master equation can be written in matrix form.
Transforming to a new basis $\vec{S}\,\PI = (p_{0},p_{1}/2+\vec{I}\cdot\vec{n}_{L},p_{2},p_{1}/2-\vec{I}\cdot\vec{n}_{L},\vec{I}\cdot(\vec{n}_{z}\times\vec{n}_{L}),I_{z})$ similar to the above, the kernel of the spinless system $\vec{W}_\text{sl}$ again assumes block-diagonal form for $\phi=2\pi m$:
\begin{widetext}
\begin{equation}	\label{eq:Wsl}
	\vec{S}\vec{W}_\text{sl}\vec{S}^{-1}  = 	
	\left(\begin{array}{cc|cc|cc}
		-2\GL		& G_R^+			& 0		& G_R^-			& \cf 2\GR\si	& 0	\\
		G_L^+		& -G_L^--G_R^+		& G_R^-		& 0			& \Gamma\si	& 0	\\\hline
		0		& G_L^-			& -2\GR		& G_L^+			& \cf 2\GR\si	& 0	\\
		G_L^-		& 0			& G_R^+		& -G_L^+-G_R^-		& -\Gamma\si	& 0	\\\hline
		-\cf \GL\si	& -\frac{1}{2}\Gamma\si	& -\cf\GR\si	& -\frac{1}{2}\Gamma\si	& -\GL-\GR	& 0	\\
		0		& 0			& 0		& 0			& 0		& -\GL-\GR
	\end{array}\right)  ,
\end{equation}
\end{widetext}
where, as above, $G_{r}^{\pm} = \cf \Gamma_{r}(1\pm\co)$ was defined.

The block structure can be understood by realizing that, for an AB-phase $\phi=2\pi m$, charging the empty double dot $\bra{0}$ from the source always results in the symmetric state $\bra{+}=(\bra{1,0}+\bra{0,1})/\sqrt{2}$. From this state the electron may leave to the drain, resulting again in $\bra{0}$. In contrast to the spinful case, the symmetric state $\bra{+}$ cannot be charged with a second electron, so that $\bra{0}$ and $\bra{+}$ constitute a decoupled set of states, whose motion is described by the upper left block of Eq.~\eqref{eq:Wsl}.
On the other hand, the doubly occupied state $\bra{1,1}$ may loose one electron to the drain, resulting in the antisymmetric combination $\bra{-}=(\bra{1,0}-\bra{0,1})/\sqrt{2}$, which can also be charged again from the source, but cannot be discharged to the drain. The two states $\{\bra{-},\bra{1,1}\}$ therefore also form a decoupled set, which is described by the middle block of Eq.~\eqref{eq:Wsl}. The remaining components of the isospin are unoccupied.

\begin{figure}[b]
	\includegraphics[width=.75\columnwidth]{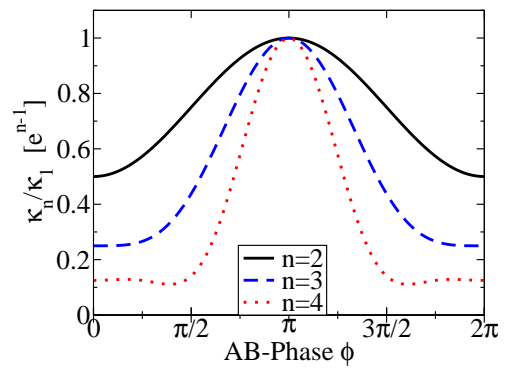}
	\caption{The normalized $n$th cumulants $\kappa_{n}/\kappa_{1}$ show no divergence for $\phi=2\pi m$ if spin is neglected. Instead they assume the values expected for a non-interacting two-level system. At $\phi=(2m+1)\pi$ the statistics becomes Poissonian due to an isospin blockade effect.}
	\label{fig:spinless}
\end{figure}

In contrast to the situation with spinful electrons the statistics of the two subspaces are the same, regardless of the coupling strengths. This is owed to the fact that both subspaces are two-dimensional and describe a single level, the statistics of which is symmetric in source and drain coupling. Correspondingly the statistics of the spinless model becomes that of two independent, non-interacting levels\cite{jong,bagrets:2003} for $\phi=2\pi m$ (see Fig.~\ref{fig:spinless}).

The figure also shows that at $\phi=(2m+1)\pi$ the statistics becomes Poissonian. This is due to an isospin-blockade: Adding one electron to the empty system results in the state $(\bra{1,0}-i\bra{0,1})/\sqrt{2}$. This state cannot decay to the drain. On the other hand, the doubly occupied state may loose one electron, resulting in the combination $(\bra{1,0}+i\bra{0,1})/\sqrt{2}$, which in turn cannot be refilled from the source. As a consequence, the system is trapped in the singly occupied state and transport events become increasingly rare as the flux approaches $2\pi m$, resulting in Poissonian statistics.

In summary, no super-Poissonian noise is predicted for any value of the magnetic flux when spin is neglected.
In Ref.~\onlinecite{gurvitz} diverging noise was also reported for spinless electrons as the flux approaches multiples of the flux quantum, $\phi=2\pi m$. There a double dot Aharonov-Bohm interferometer with non-degenerate energy levels (level splitting $\Delta\epsilon$) was considered, with strong intra-dot interaction, so that double occupation is forbidden and a charge blockade mechanism can interrupt transport.
It was pointed out that it is then important to perform the limit correctly as the system approaches the degeneracy point, i.e., taking first $\phi\rightarrow 0$ and then $\Delta\epsilon\rightarrow 0$.
Since we do not assume Coulomb interaction between the dots a charge blockade mechanism is not effective and the order of limits is uncritical.

The spinless double-dot system with degenerate levels discussed here can be mapped onto a non-interacting quantum-dot spin-valve\cite{braun:2004}, with perfect lead polarization. The statistics of quantum-dot spin-valves have been analyzed in more general contexts, both without and with Coulomb interaction.\cite{lindebaum:2009}

\section{Conclusion}    \label{sec:conclusion}

We have analyzed the full counting statistics of electronic transport through an Aharonov-Bohm interferometer with two quantum dots embedded in its arms in the shot-noise regime.
We found that for values of the Aharonov-Bohm flux that are integer multiples of the flux quantum, 
the second and higher cumulants diverge.
This divergence is related to a separation of the Hilbert space of the double dot into disconnected subspaces that contain the spin singlet and triplet state for double occupancy, respectively.
As the two subspaces have different transport statistics, the system exhibits electron bunching that results in strongly super-Poissonian statistics.
The coupling between the two subspaces, and therefore also the appearance of the divergence, is tunable by the Aharonov-Bohm flux.
The inclusion of spin in the description of the system has been shown to be crucial for the occurrence of diverging moments.
Furthermore, we discussed how the divergence is cut off by the influence of an environment that relaxes either the isospin or the spin coherence.

\acknowledgments

We acknowledge useful discussions with R.~Fazio, G.~Kie{\ss}lich and financial support from DFG via SPP~1285.

\appendix

\section{Perturbative expansion for lower order cumulants}

\newcommand{\Wb}{\overline{\vec{W}}}
As mentioned in Sec.~\ref{sec:results} we make use of a recursive scheme to calculate the cumulants based on a master equation, which was developed by Flindt {\em et al.}.\cite{flindt:2008} It is based on the assumption that the system's behavior is dominated by a single pole of the kernel $\vec{W}(\chi) = \vec{W}(0)+\Wb(\chi)$. The cumulant generating function is then found to be
\begin{equation}\label{eq:flindt}
	S(\chi)	=	\ket{0}	\Wb(\chi) \left[ 1+\vec{R}(\chi) \Wb(\chi) \right]^{-1}	\bra{0}	,
\end{equation}
where $\ket{0}$ and $\bra{0}$ are the left and right nullvectors of $\vec{W}(0)$ and $\vec{R}(\chi)$ is the pseudoinverse $\vec{R}(\chi) = \vec{Q} \left[ \vec{W}(\chi)-S(\chi) \right]^{-1} \vec{Q}$, with $Q=\vec{1}-\bra{0}\ket{0}$.

The cumulants can be obtained from Eq.~(\ref{eq:flindt}) using Brillouin-Wigner perturbation theory in $\chi$.
This requires knowledge of the derivatives of the kernel.
The first two cumulants (current and noise) are:
\begin{eqnarray}
	\kappa_{1}	&=&	\frac{1}{i} 		\ket{0}	\Wb' 							\bra{0}	\\
	\kappa_{2}	&=&	\frac{1}{i^2}	\ket{0}	\Wb''  -  2\Wb' \vec{R} \Wb'			\bra{0}	,
\end{eqnarray}
where the prime denotes a derivative w.r.t.~the counting field~$\chi$.

The multiplication with left and right eigenvectors automatically picks out the correct eigenvalue.
In our system we find the left nullvector to be $\ket{0} = (1,1,1,1,0,0,0) $, so that it also automatically takes care of getting rid of the unneeded off-diagonal elements mentioned above.


\begin{thebibliography}{99}


    \bibitem{levitov:1996}
	L.~S.~Levitov and G.~B.~Lesovik,
	Pis'ma Zh.~Eksp.~Teor.~Fiz.~{\bf 58}, 225 (1993)
	[JETP Lett.~{\bf 58}, 230 (1993)];
	L.~S.~Levitov, H.-W.~Lee, and G.~B.~Lesovik,
	J.~Math.~Phys.~{\bf 37}, 4845 (1996).
    \bibitem{QNmesPhys}
	A good overview is found in {\em Quantum Noise in Mesoscopic Physics}, NATO Science Series,
	edited by Yu.~V.~Nazarov (Kluwer, Dordrecht, 2003).
    \bibitem{nazarov:2002}
        Yu.~V.~Nazarov and D.~A.~Bagrets,
        Phys.~Rev.~Lett.~{\bf 88}, 196801 (2002).
    \bibitem{KeldyshGF}
        W.~Belzig and Yu.~V.~Nazarov,
        Phys.~Rev.~Lett.~{\bf 87}, 067006 (2001);
        Yu.~V.~Nazarov and M.~Kindermann,
        Eur.~Phys.~J.~B~{\bf 35}, 413 (2003).
    \bibitem{stochasticPathint}
        S.~Pilgram, A.~N.~Jordan, E.~V.~Sukhorukov, and M.~B\"uttiker,
        Phys.~Rev.~Lett.~{\bf 90}, 206801 (2003);
        A.~N.~Jordan, E.~V.~Sukhorukov, and S.~Pilgram,
        J.~Math.~Phys.~{\bf 45}, 4386 (2004).
    \bibitem{emary:2007}
        C.~Emary, D.~Marcos, R.~Aguado, and T.~Brandes,
        Phys.~Rev.~B {\bf 76}, 161404(R) (2007).

    \bibitem{TunnelDots}
        D.~A.~Bagrets, Y.~Utsumi, D.~S.~Golubev, and G.~Sch\"on,
        Fortschritte der Physik {\bf 54}, 917 (2006).
    \bibitem{bagrets:2003}
        D.~A.~Bagrets and Y.~V.~Nazarov,
        Phys.~Rev.~B {\bf 67}, 085316 (2003).
    \bibitem{kiesslich:2006}
        G.~Kie{\ss}lich, P.~Samuelsson, A.~Wacker, and E.~Sch\"oll,
        Phys.~Rev.~B {\bf 73}, 033312 (2006).
    \bibitem{gogolin:2006}
        A.~O.~Gogolin and A.~Komnik,
        Phys.~Rev.~Lett.~{\bf 97}, 016602 (2006).
	\bibitem{emary:2009}
		C.~Emary,
		arXiv:0902.3544v1 [cond-mat.mes-hall].
    \bibitem{braggio:2006}
        A.~Braggio, J.~K\"onig, and R.~Fazio,
        Phys.~Rev.~Lett.~{\bf 96}, 026805 (2006).
    \bibitem{flindt:2008}
        C.~Flindt, T.~Novotn\'y, A.~Braggio, M.~Sassetti, and A.-P.~Jauho,
        Phys.~Rev.~Lett.~{\bf 100}, 150601 (2008).

    \bibitem{fricke:2007}
        C.~Fricke, F.~Hohls, W.~Wegscheider, and R.~J.~Haug,
        Phys.~Rev.~B~{\bf 76}, 155307 (2007).
    \bibitem{fujisawa:2006}
        T.~Fujisawa, T.~Hayashi, R.~Tomita, and Y.~Hirayama,
        Science~{\bf 312}, 1634 (2006).
    \bibitem{gustavsson:2006}  
        S.~Gustavsson, R.~Leturcq, B.~Simovic, R.~Schleser, P.~Studerus, T.~Ihn, K.~Ensslin, D.~C.~Driscoll, and A.~C.~Gossard,
        Phys.~Rev.~B {\bf 74}, 195305 (2006).
    \bibitem{gustavsson:interference}
	S.~Gustavsson, M.~Studer, R.~Leturcq, T.~Ihn, K.~Ensslin, D.~C.~Driscoll, and A.~C.~Gossard,
	Phys.~Rev.~B {\bf 78}, 155309 (2008);
	S.~Gustavsson, R.~Leturcq, M.~Studer, T.~Ihn, K.~Ensslin, D.~C.~Driscoll, and A.~C.~Gossard,
	Nano Lett.~{\bf 8}, 2547 (2008).


    \bibitem{qubitdetector}
        R.~J.~Schoelkopf, A.~A.~Clerk, S.~M.~Girvin, K.~W.~Lehnert, and M.~H.~Devoret,
        in {\it Quantum Noise in Mesoscopic Physics}, edited by Yu.~V.~Nazarov (Kluwer, Dordrecht, 2003).
    \bibitem{thermalescape}
        B.~Huard, H.~Pothier, N.~O.~Birge, D.~Esteve, X.~Waintal, and J.~Ankerhold,
        Ann.~Phys.~{\bf 16}, 736 (2007).
    \bibitem{cooperpairtunneling}
        R.~K.~Lindell, J.~Delahaye, M.~A.~Sillanp\"a\"a, T.~T.~Heikkil\"a, E.~B.~Sonin, and P.~J.~Hakonen,
        Phys.~Rev.~Lett.~{\bf 93}, 197002 (2004);
        T.~T.~Heikkil\"a, P.~Virtanen, G.~Johansson, and F.~K.~Wilhelm,
        ibid.~247005 (2004);
        A.~V.~Timofeev, M.~Meschke, J.~T.~Peltonen, T.~T.~Heikkil\"a, and J.~P.~Pekola,
        Phys.~Rev.~Lett.~{\bf 98}, 207001 (2007).
    \bibitem{MQTfourthcumulant}     
        J.~Ankerhold and H.~Grabert,
        Phys.~Rev.~Lett.~{\bf 95}, 186601 (2005).
    \bibitem{thresholddetectors}
        J.~Tobiska and Yu.~V.~Nazarov,
        Phys.~Rev.~Lett.~{\bf 93}, 106801 (2004);
        J.~P.~Pekola,
        Phys.~Rev.~Lett.~{\bf 93}, 206601 (2004).

	\bibitem{cottet:2004}
		A.~Cottet, W.~Belzig, and C.~Bruder,
		Phys.~Rev.~B {\bf 70}, 115315 (2004).
	\bibitem{kiesslich:2007}
		G.~Kie{\ss}lich, E.~Sch\"oll, T.~Brandes, F.~Hohls, and R.~J.~Haug,
		Phys.~Rev.~Lett.~{\bf 99}, 206602 (2007).
	\bibitem{sanchez:2008}
		R.~S\'anchez, S.~Kohler, P.~H\"anggi, and G.~Platero,
		Phys.~Rev.~B {\bf 77}, 035409 (2008).
		
	\bibitem{aghassi}
		J.~Aghasssi, A.~Thielmann, M.~H.~Hettler and G.~Sch\"on,
		Appl.~Phys.~Lett.~{\bf 89}, 052101 (2006);
		J.~Aghassi, A.~Thielmann, M.~H.~Hettler, and G.~Sch\"on,
		Phys.~Rev.~B {\bf 73}, 195323 (2006).

	\bibitem{bistability}
		O.~A.~Tretiakov and K.~A.~Matveev,
		Phys.~Rev.~B {\bf 71}, 165326 (2005).
	\bibitem{jordan:2004}
		A.~N.~Jordan and E.~V.~Sukhorukov,
		Phys.~Rev.~Lett.~{\bf 93}, 260604 (2004).
	\bibitem{flindt:2005}
		C.~Flindt, T.~Novotn\'y, and A.-P.~Jauho,
		Europhys.~Lett.~{\bf 69}, 475 (2005).

	\bibitem{belzig:2005}
		W.~Belzig,
		Phys.~Rev.~B {\bf 71}, 161301(R).
	\bibitem{bulka}
		B.~R.~Bu{\l}ka,
		Phys.~Rev.~B {\bf 62}, 1186 (2000).
\bibitem{braun:noise}
	M.~Braun, J.~K\"onig, and J.~Martinek,
	Phys.~Rev.~B {\bf 74}, 075328 (2006).
	\bibitem{safonov}
		S.~S.~Safonov, A.~K.~Savchenko, D.~A.~Bagrets, O.~N.~Jouravlev, Y.~V.~Nazarov, E.~H.~Linfield, and D.~A.~Ritchie,
		Phys.~Rev.~Lett.~{\bf 91}, 136801 (2003).
	\bibitem{djuric:2005}
		I.~Djuric, B.~Dong, and H.~L.~Cui,
		Appl.~Phys.~Lett.~{\bf 87}, 032105 (2005).
	\bibitem{kiesslich:2003}
		G.~Kie{\ss}lich, A.~Wacker, and E.~Sch{\"o}ll,
		Phys.~Rev.~B {\bf 68}, 125320 (2003).
\bibitem{gurvitz}
	F.~Li, H.~Jiao, J.~Luo, X.-Q.~Li, S.A.~Gurvitz,
	arXiv:0812.0846v1;
	Feng Li, Xin-Qi Li, Wei-Min Zhang, S.A.~Gurvitz,
	arXiv:0803.1618v1.
	\bibitem{sanchez:2007}
		R.~S\'anchez, G.~Platero, and T.~Brandes,
		Phys.~Rev.~Lett.~{\bf 98}, 146805 (2007);
		Phys.~Rev.~B {\bf 78}, 125308 (2008).


	\bibitem{weymann:2008}
		I.~Weymann,
		Phys.~Rev.~B {\bf 78}, 045310 (2008).

	\bibitem{spinless2} 
		S.-K.~Wang, H.~Jiao, F.~Li, X.-Q.~Li, and Y.~J.~Yan,
		Phys.~Rev.~B~{\bf 76}, 125416 (2007).
\bibitem{guo:2008}
	L.~Qin and Y.~Guo,
	J.~Phys.:~Condens.~Matter {\bf 20}, 365206 (2008).

\bibitem{bodoky:2008}
	F.~Bodoky, W.~Belzig, and C.~Bruder,
	Phys.~Rev.~B~{\bf 77}, 035302 (2008).
	\bibitem{spinless1}
		B.~Dong, X.-L.~Lei, and N.J.M.~Horing,
		J.~Appl.~Phys.~{\bf 104}, 033532 (2008);
		Phys.~Rev.~B {\bf 77}, 085309 (2008);
		B.~Dong, X.-L.~Lei, and H.-L.~Cui,
		Commun.~Theor.~Phys.~{\bf 49}, 1045 (2008).


	\bibitem{burkhard:1999}
		G.~Burkard, D.~Loss, and D.~P.~DiVincenzo,
		Phys.~Rev.~B {\bf 59}, 2070 (1999).
	\bibitem{loss:2000}
		D.~Loss and E.~V.~Sukhorukov,
		Phys.~Rev.~Lett.~{\bf 84}, 1035 (2000).

	\bibitem{legel:2007}
		S.~Legel, J.~K\"onig, G.~Burkard, and G.~Sch\"on,
		Phys.~Rev.~B~{\bf 76}, 085335 (2007).
	\bibitem{legel:2008}
		S.~Legel, J.~K\"onig, and G.~Sch\"on,
	 	New.~J.~Phys.~{\bf 10}, 045016 (2008).

	\bibitem{michaelis}
		B.~Michaelis, C.~Emary, and C.~W.~J.~Beenakker,
		Europhys.~Lett.~{\bf 73}, 677 (2006).
	\bibitem{xu}
		X.~Xu, B.~Sun, P.~R.~Berman, D.~G.~Steel, A.~S.~Bracker, D.~Gammon, and L.~J.~Sham,
	Nature Physics {\bf 4}, 692 (2008).

    \bibitem{diagrams}
        J.~K\"onig, H.~Schoeller, and G.~Sch\"on,
        Phys.~Rev.~Lett.~{\bf 76}, 1715 (1996);
        J.~K\"onig, J.~Schmid, H.~Schoeller, and G.~Sch\"on,
        Phys.~Rev.~B {\bf 54}, 16820 (1996).
    \bibitem{technique}
        H.~Schoeller, in {\it Mesoscopic Electron Transport}, edited by L.~L.~Sohn, L.~P.~Kouwenhoven, and G.~Sch\"on (Kluwer, Dordrecht, 1997);
        J.~K\"onig, {\it Quantum Fluctuations in the Single-Electron Transistor} (Shaker, Aachen, 1999).

\bibitem{urban:2008}
	D.~Urban, J.~K\"onig, and R.~Fazio,
	Phys.~Rev.~B {\bf 78}, 075318 (2008).


\bibitem{note-1}
The divergence can be understood from a minimal model describing two states, both carrying a Poissonian current with different rates $\Gamma_1\neq\Gamma_2$. If the system switches between the states with a rate $\Omega$ the kernel of the system's master equation reads
$\vec{W}=\left(\begin{array}{cc}
	\Gamma_{1}(e^{i\chi}-1)	& \Omega \\
	\Omega					& \Gamma_{2}(e^{i\chi}-1)
\end{array}\right)$.
The cumulant generating function is readily found to be $S(\chi)=\left( (\Gamma_{1}+\Gamma_{2})(e^{i\chi}-1)-2\Omega+\sqrt{(\Gamma_{1}-\Gamma_{2})^{2}(e^{i\chi}-1)+4\Omega^{2}} \right)/2$. The normalized moments diverge as $\Omega\rightarrow0$, the second in particular reads $\kappa_{2}/\kappa_{1}=1+(\Gamma_{1}-\Gamma_{2})^2/(2\Omega(\Gamma_{1}+\Gamma_{2}))$.


\bibitem{hassler:2008}
	F.~Hassler, M.~V.~Suslov, G.~M.~Graf, M.~V.~Lebedev, G.~B.~Lesovik, and G.~Blatter,
	Phys.~Rev.~B {\bf 78}, 165330 (2008).
\bibitem{taddei:2002}
	F.~Taddei and R.~Fazio,
	Phys.~Rev.~B {\bf 65}, 075317 (2002).

\bibitem{urban:sftrans}
	D.~Urban, M.~Braun, and J.~K\"onig,
	Phys.~Rev.~B {\bf 76}, 125306 (2007).

\bibitem{jong}
	M.~J.~M.~de Jong,
	Phys.~Rev.~B {\bf 54}, 8144 (1996).

\bibitem{braun:2004}
		M.~Braun, J.~K\"onig, and J.~Martinek,
		Phys.~Rev.~B {\bf 70}, 195345 (2004).

\bibitem{lindebaum:2009}
	S.~Lindebaum, D.~Urban, J.~K\"onig,
	arXiv:0903.1759.

\end{thebibliography}
\end{document}